\documentclass[usenatbib]{mn2e}
\usepackage{psfig}
\title[HST Observations of SV Cam]{Hubble Space Telescope Observations of SV Cam: I. The Importance of Unresolved Starspot Distributions in Lightcurve Fitting}

\author[S.V. Jeffers et al.]
       {S.V.Jeffers$^{1,2}$, J.R.Barnes$^2$, A.Collier Cameron$^2$,
 J.-F. Donati$^1$ \\
$^1$ Laboratoire d'Astrophysique, Observatoire Midi-Pyr$\acute{e}$n$\acute{e}$es, 14, avenue Edouard Belin, F-31400 Toulouse, France \\
$^2$ School of Physics and Astronomy, University of St Andrews, North Haugh, St Andrews, Fife KY16 9SS, U.K.\\ }

\date{}

\pagerange{\pageref{firstpage}--\pageref{lastpage}}
\pubyear{2005}

\begin{document}

\maketitle

\label{firstpage}

\begin{abstract}

We have used maximum entropy eclipse mapping to recover images of the
visual surface brightness distribution of the primary component of the
RS CVn eclipsing binary SV Cam, using high-precision photometry data
obtained during three primary eclipses with STIS aboard the Hubble
Space Telescope.  These were augmented by contemporaneous ground-based
photometry secured around the rest of the orbit. The goal of these
observations was to determine the filling factor and size distribution
of starspots too small to be resolved by Doppler imaging.  The
information content of the final image and the fit to the data were
optimised with respect to various system parameters using the
$\chi^{2}$ landscape method, using an eclipse mapping code that solves
for large-scale spot coverage.  It is only with the unprecedented
photometric precision of the HST data (0.00015 mag) that it is
possible to see strong discontinuities at the four contact points in the
residuals of the fit to the lightcurve.  These features can only be
removed from the residual lightcurve by the reduction of the
photospheric temperature, to synthesise high unresolvable spot
coverage, and the inclusion of a polar spot.  We show that this
spottedness of the stellar surface can have a significant impact on
the determination of the stellar binary parameters and the fit to the
lightcurve by reducing the secondary radius from
0.794$\pm$0.009\,R$_\odot$ to 0.727$\pm$0.009\,R$_\odot$.  This new
technique can also be applied to other binary systems with high
precision spectrophotometric observations.

\end{abstract}


\section{Introduction}

%RS CVn binary systems  magnetic activity 

Star-spots and other forms of magnetic activity are prevalent on
rapidly rotating stars with temperatures low enough to have outer
convective zones.  The dynamo activity in these rapidly rotating cool
stars leads to the suppression of convection over large areas of the
stellar photosphere.  These giant starspots can modulate the light of
such active stars as they rotate by up to tens of percent.  In
general, stars that exhibit this type of activity are solar-type stars
with rotational periods of less than one day.  This rapid rotation is
seen in half the G and K dwarfs in open clusters younger than
100\,Myr, and may also persist into middle age through tidal locking
in a close binary systems such as RS CVns.

Doppler imaging is a powerful tool for determining where spots
congregate on the stellar surface, but it is less successful at
determining how much overall spot activity is present.  Other methods
such as TiO-band monitoring studies \citep{oneal98tio} indicate that
between 30-50\% of the stellar surface is spotted compared to 10-15\%
from Doppler imaging \citep{jeffers02}.  The discrepancy in these
methods can be accounted for if the stellar surface is heavily covered
by spots that are too small to be resolved through Doppler imaging. To
try to resolve this we have used the photometric capabilities of the
HST (signal-to-noise $\simeq$ 5000) to eclipse-map the inner face of the F9V
primary of the eclipsing binary SV Cam.  Using these observations,
\citet{jefferspc05} found that the surface flux in the low-latitude
region to be approximately 30\% lower than the best fitting {\sc
phoenix} model atmosphere.  A possible cause of this flux deficit is
that the primary star's surface is peppered with unresolvable spots.
They also found that when the 30\% spottedness is extended to the
entire stellar surface there still remains a flux deficit, which could
be explained by the presence of a large polar cap.

%paragraph about lightcurve fitting and how nobody ever includes 
%polar caps etc... have a look at the papers on the GAIA photo of 
%Alexis.

In this paper we use the Maximum Entropy eclipse-mapping code DoTS
(\citealt{cameron97dots}; \citealt{cameron97xyuma}) to firstly solve
for the system parameters of the lightcurve, and then reconstruct the
surface brightness distribution of SV Cam's primary star.  We also
show how a synthetic polar cap and the star being peppered with spots
below the resolution limits of eclipse mapping are essential parameters
for obtaining a good lightcurve fit.

\begin{table*}

\begin{tabular}{c c c c c c c c c}
%\begin{tabular}{ l l l l l l } 

\hline
\hline

{Visit} & {Obs. Date} & {Frame No} &  {UT Start} & {UT End} & {Julian Date (+ 52210)} & {Exposure Time (s)} & {Phase Range} \\

\hline

1 & 01 Nov 01 & 1-48 & 21:04:33 & 21:47:44 & 4.879394-4.908758 & 30 & 0.901-0.950 \\ 
& & 49-107 & 22:24:18 & 23:17:41  &  4.909383-4.971851 & 30  & 0.952-1.058 \\
& & 108-165 & 00:00:23  & 00:51:58 & 5.001505-5.037328  & 30 & 1.107-1.168  \\
2 & 03 Nov 01  & 166-213  & 14:43:36  & 15:26:47  & 6.614915-6.644280  & 30 & 0.828-0.877 \\
& & 214-272 & 16:02:23  & 16:55:46  &  6.669628-6.706701  & 30 & 0.920-0.983 \\
& & 273-330 & 17:38:27  & 18:30:02 & 6.736343-6.772167  & 30 & 1.033-1.093 \\
3 & 05 Nov 01 & 331-378 & 14:43:36 & 15:26:47 & 8.416575-8.446561  & 30 & 0.865-0.916 \\
& & 379-437 & 16:02:23 & 16:55:46  & 8.470752-8.507825  & 30 & 0.957-1.019 \\
& & 438-495 & 17:38:27  & 18:30:02 & 8.537468-8.573291  & 30 & 1.069-1.13 \\

\hline
\hline
\end{tabular}
\caption{Journal of observations for SV Cam using the Space Telescope Imaging Spectrograph on board the Hubble Space Telescope}
\label{obs2}
\end{table*}

\section{Observations}

SV Cam (F9V + K4V, P=0.59d, vsini=102 km\,s$^{-1}$, $i$=90$^\circ$) is
a totally eclipsing, short-period binary system originally identified
by \citet{guthnick29}.  Contemporaneous HST and ground based
observations were obtained to eclipse map SV Cam. The ground based
photometric observations were obtained to complete the light curve
outside primary eclipse making it possible to determine any global
asymmetry in the lightcurve due to the presence of any uneclipsed
spots.

\subsection{HST Observations}

%also have a look at Brown Charbonneau planet paper for more of an idea
%as to how to sell the Hubble data.

\begin{figure}
%\hspace{0.5cm}
\psfig{file=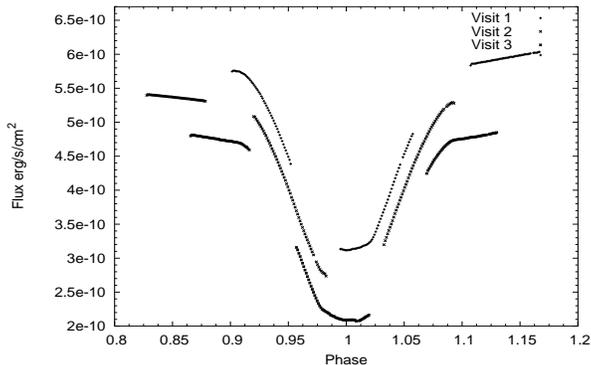,width=8cm,height=5cm,angle=270}
\caption{The comprising sections of the primary eclipse from each of
the 3 HST visits, with an off-set of 5$\times$10$^{-11}$ ergs/s/cm$^2$
for clarity.}
\label{visit}
\end{figure}

We observed three primary eclipses of SV Cam during three HST visits
at intervals of 2 days between 2001 November 1 to 5.  A total of 9
spacecraft orbits were devoted to the observations.  The G430L grating
of Space Telescope Imaging Spectrograph (STIS) was used to disperse
starlight over 2048 CCD elements. The observations are summarised in
Table~\ref{obs2} and the lightcurve for each visit is shown in 
Fig.~\ref{visit}.

The observations cover a wavelength range from 2900\,\AA\,\, to
5700\,\AA, with an exposure time of 30\,s and a cadence of 40\,s.  The
exposure time was chosen as a trade-off between maximising
signal-to-noise per exposure and minimising the time interval between
observations.  The total count over the entire detector is
9.4$\times$10$^7$ electrons per exposure, with the photometric precision
calculated to be 0.00015 magnitudes (S:N = 5000).  The main
source of systematic error, due to the subtle motion of the satellite,
was removed by cross-correlating each frame with respect to a template
spectrum, and then adding or subtracting a scaled number of counts.
After correcting for this `jitter' the RMS scatter was consistent with
the error expected from photon noise.

\subsection{Ground Based Photometry}

The ground based photometric observations were obtained with the
0.93-m James Gregory Telescope (JGT) at the University of St Andrews,
using a Wright Instruments CCD camera mounted at the Cassegrain
focus. The observations were made with a narrow-band filter whose
85\,\AA\ (FWHM) bandpass is centred at 5300\,\AA, close to the V band.
Using an exposure time of 60\,s we obtained a photometric precision of
$\pm 0.006$\,mag in good photometric conditions.  The overhead for the
chip readout between 60\,s observations is typically 15\,s, giving an
overall sampling interval of 75\,s.  Observations were made over a two
week period centred at the time of the HST observations.  The weather
permitted the observation of four full primary eclipses and three full
secondary eclipses over this period.  

%The field of view containing the
%target and reference stars is shown in Fig.~\ref{fchart}.

%\begin{figure}
%\hspace{1cm} \psfig{file=svcam176a.ps,width=6.2cm,height=4.2cm,angle=0}
%\caption{Finding chart showing SV Cam and comparison star positions.  
%Differential magnitudes were computed with respect to the
%first comparison star, while the second star served as a check.  The
%field size is approximately 11$\times$17\,arcmin$^2$.  }
%\label{fchart}
%\end{figure}

\section{Data Processing}

\subsection{Data Reduction}

Reduction of the HST data comprised calibrating the two dimensional
CCD images and extracting one dimensional spectra using the standard
STIS data reduction pipeline.  A detailed description of this
procedure is contained in the STIS handbook.  This pipeline routine
produces bias and dark subtracted, flat-fielded wavelength calibrated
images.  The auto-wavecal was disabled for these images, but instead
two `Guest Observer' wavecals were taken per orbit.  Cosmic rays were
removed using an algorithm that identifies and rejects cosmic rays and
other non repeatable defects by comparing successive frames.

The ground-based photometric data were reduced using {\sc jgtphot}, a
software package developed for use with the James Gregory Telescope at
St Andrews \citep{bell93jgt}.  The resulting light curves are calculated
as differential magnitude values with respect to the marked comparison
and check stars.

\subsection{Time Correction of HST data set}

As the time of the extracted one-dimensional spectra that is output
from the standard STIS data reduction pipeline (calstis) is in UT, it
is necessary to apply a further time correction to this value.  The
heliocentric correction due to the motion of the Earth around the
Sun, and due to the orbital motion of the satellite was determined
through use of the `odelaytime' routine.  To convert UT into atomic
time 32 leap seconds were added, and to convert atomic time into
terrestrial time 32.184 seconds were added, resulting in the correct
terrestrial time value.  Absolute precision in HST time is required as
it is necessary to have the primary eclipse of the HST and of the
ground-based observations exactly aligned.

\subsection{Interpolation of Data Sets}

In order to analyse the absolute spectrophotometry from STIS together
with the ground-based differential photometry, it was necessary to
place the magnitudes derived from the two instruments on a common
magnitude scale.  We constructed a synthetic filter, V$_{hst}$, with a
bandpass and effective wavelength matching those of the filter used
for the JGT observations, (bandpass of 80\AA\ and effective wavelength
5300\AA) and applied it to the HST observations. Similarly we
constructed a digital version of the Johnson B filter, B$_{hst}$.

\begin{figure}
\psfig{file=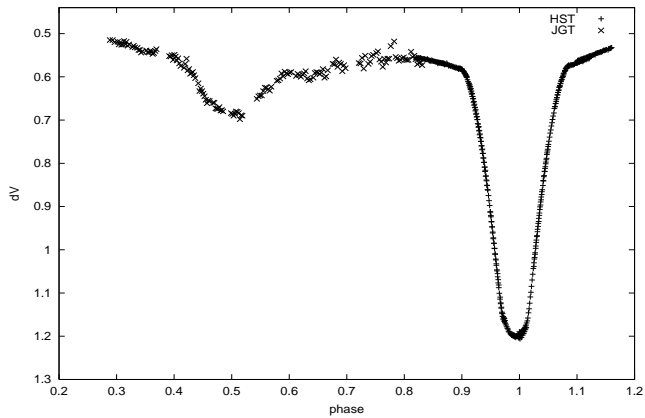,width=8.7cm,height=5.6cm,angle=270}
\caption{Interpolated HST data (primary eclipse) set plotted with JGT
data set (secondary eclipse).}
\label{interpol}
\end{figure}

At the time of each JGT flux measurement during primary eclipse we
interpolated the V$_{hst}$ magnitude and the ($B_{hst}-V_{hst}$)
colour index from the two nearest points bracketing the same phase.
%\begin{figure}
%\psfig{file=colour.ps,width=8.7cm,height=5.6cm,angle=270}
%\caption{Colour-colour plot for interpolation of the HST and
%ground-based data sets}
%\label{colourplot}
%\end{figure}
The colour equation relating the two magnitude systems 
was then derived by a linear least-squares fit 
to the plot of $\delta V _{jgt} - V_{hst}$ versus
($B_{hst}-V_{hst}$), as shown in Fig.~\ref{interpol}. This yielded
the colour equation
\begin{equation}
    \delta V _{jgt} - V_{hst} = \alpha(B_{hst}-V_{hst}) + \beta,
    \label{eq:coloureq}
\end{equation}
where the coefficient of the colour term $\alpha = -0.0725392 \pm 0.05836$ 
and the zero-point $\beta = -8.69367 \pm 0.03318$ magnitudes.  
%The fit is shown in Fig.~\ref{colourplot}.

Using this linear calibration, all of the V$_{hst}$ magnitudes were
converted to the JGT system.  This procedure takes account of the zero
point off-set, the effect of the V$_{hst}$ and the JGT observations
having slightly different effective wavelengths, and the different
spectral responses of the two filter sets and the two CCDs.  The
interpolated HST data set is shown in Fig.~\ref{interpol} along with
the JGT data.

\begin{table*}
\fontsize{7}{9}\selectfont  

\begin{tabular}{ l c c r@{.}l r@{.}l r@{.}l r@{.}l  l l }
\hline
\hline

Ref. & T$_{pri}$ & T$_{sec}$ 
&\multicolumn{2}{c}{R$_{pri}$} 
&\multicolumn{2}{c}{R$_{sec}$} 
&\multicolumn{2}{c}{{\em i}} 
&\multicolumn{2}{c}{q} 
& S\,/\,P 
& Spectral Type \\
\hline
This work & 6038(58) & 4804(143) & 1 & 238(6) & 0 & 794(6)   & \multicolumn{2}{c}{} & \multicolumn{2}{c}{}& P & F9V+K4V  \\

(reduced T$_{phot}$ + polar cap) & 5935(38) & 4808(143) & 1 & 235(7) & 0 & 727(6)   & \multicolumn{2}{c}{} & \multicolumn{2}{c}{}& P &  G0V+K4V \\

\citet{lehmann02} & &  & 1 & 18(2) & 0 & 76(2) & 90 & 0 & 0 & 642(5) & S & G0V+K6V\\

\citet{rucinski02} & &\multicolumn{2}{c}{} & \multicolumn{2}{c}{}& & 90 & 0 & 0 & 641(7) & S & G2V(pri+sec) \\

\citet{kjurkchieva02} &  & & 1 & 38(5) & 0 & 94(6) & 80 & 0 & 0 & 593(11) & S & F5V+G8V\\

\citet{albayrak01} & 6440 & 4467(34) & 1 & 38(2) & 0 & 87(2) & 89 & 6(9) & 0 & 56 & P & F8V+K6V\\

\citet{albayrak01} & 6200 & 4377(32) & 1 & 38(2) & 0 & 87(2) & 89 & 6(9) & 0 & 56(10) & P & F5V+K6V\\

\citet{pojmanski98} &  &  & 1 & 25 & 0 & 8 & 90 & 0 & 0 & 56(9)& S & F5V+K0V\\

\citet{patkos94} & 5750 & 4500 & 1 & 18(5) & 0 & 75(3) &\multicolumn{2}{c}{}  &\multicolumn{2}{c}{} & P & G2/3V+K4/5  \\

\citet{rainger91xyuma} & &  &\multicolumn{2}{c}{} &\multicolumn{2}{c}{} & 90 &0 & 0 & 7 & S & G2/3V+K4V    \\

\citet{zelik88} & 5800 & 4300 & 1 & 17(3) & 0 & 74(3) & 89 & 5(5) & \multicolumn{2}{c}{} &  P & G3V\\

\citet{budding87} & 5750 & 4500 & 1 & 11(2) & 0 & 74(2) & 90 & 0(5) & 0 & 71 & P & G3V+K4 \\

\citet{hilditch79} & 5800 & 4140 & 1 & 224 & 0 & 864 & 80 & 0 & 0 & 7 &  P & G3V+K4V\\

\hline
\hline
\end{tabular}
\caption{Summary of observed stellar parameters for SV Cam from
the literature where T$_{pri}$ and T$_{sec}$ are the primary and
secondary photospheric temperatures (K), R$_{pri}$ and R$_{sec}$ are
the primary and secondary radii (R$_\odot$), {\em i} is the binary
system inclination (degrees), q is the secondary:primary mass ratio
and S/P indicates spectroscopic or photometric
observations.}
\label{param}
\end{table*}

\section{Lightcurve Fitting}

The combined HST and JGT lightcurve of SV Cam is fitted using the maximum 
entropy code {\sc dots}.  Included in {\sc dots} is the ability to incorporate 
the surface geometry and radial velocity variations of tidally distorted 
close binary components \citep{cameron97dots} while solving for starspot 
coverage.

\subsection{Eclipse Mapping}

By using eclipse mapping it is possible to determine the detailed
locations of low latitude spots on the stellar surface.  If spots are
present on the inner surface of the primary, that is occulted by the
cooler secondary star, jagged discontinuities will be produced on the
eclipse profile as the primary is occulted by the secondary.  The
timescales and amplitudes of these discontinuities will reflect the
distribution of spot sizes on the inner surface of the primary.

\subsubsection{Maximum Entropy}

The maximum entropy method allows us to determine the simplest image
of the primary's surface (in terms of its information content as
quantified by the Shannon-Jaynes image entropy), that can reproduce,
at a specified goodness of fit, the perturbations to the primary
eclipse. 

For images where the mapping parameter $f_{i}$ is bolometric surface
brightness and so can take any positive value, the entropy takes the 
form, 
\begin{equation}
S = \sum_{i}(f_{i}-m_{i}-f_{i}\ln \frac{f_{i}}{m_{i}})
\end{equation}
Here $m_{i}$ is the default value that a pixel will have when there
are no other constraints imposed by the data.  In this analysis a
restricted form of the entropy is used as the filling-factor model has
been adopted.  This combines the entropy of the spot image $f_{i}$ and
of the photospheric image $(1-f_{i})$:
\begin{equation}
S = \sum_{i}(-f_{i}\ln \frac{f_{i}}{m_{i}}
-(1-fi)\ln \frac{(1-f_{i})}{(1-m_{i})}).
\end{equation}
  To construct the final image, the
values for the spot and photospheric $f_{i}$
are iteratively adjusted to maximise;
\begin{equation}
Q = S - \lambda \chi^{2}
\end{equation}

This is equivalent to maximising the entropy $S$ over the surface of a
hyper-ellipsoid, of constant $\chi^2$, in image space.  This is
bounded by the constraint surface at some fixed value of $\chi^{2}$.
The Lagrange multiplier, $\lambda$, is set so that the final solution
lies on a surface with $\chi^{2} \simeq M$, $M$ being the number of
measurements in the data set. By setting a Lagrangian multiplier, it
is possible to determine the extrema (maximum) of the entropy, the
image with the least information content, subject to the constraint of
obtaining the best fit to the data.  A more detailed explanation of
this, and other methods, is discussed by Collier Cameron (2001).

It is important not to over fit $\chi^2$ during the reconstruction, as
this can result in an ill-fitting photometric fit and artifacts on the
final spot map.  The number of iterations was set to 25, as through
rigorous testing we found this to be the value where the final image did not
contain any distortions due to ill fitting.  

\subsection{Geometric Parameters \label{s-param}}

The success of a surface image reconstruction depends largely on a
correct determination of the geometric parameters of the binary
system (Vincent et al 1993).  Independent reconstructions for the 
starspot parameters and the geometric parameters of the system, 
are susceptible to the distortions in the lightcurve yielding the 
wrong system parameters or vice versa.
If the system parameters are wrong, in general a satisfactory fit to
the data is only obtained by increasing the amount of structure in the
stellar surface brightness distribution, usually leading to a 
greater total spot area than is present when the correct
geometric parameters are used.

The principal parameters that affect the shape of the light-curve of
an unspotted star are the relative temperatures, the radii of the two
stars and the system inclination.  The relative surface brightnesses
of the two components determine the relative depths of the primary and
secondary eclipses.  The radii of the two stars and the binary system
inclination are related to each other by the orbital phases at which
the four contact points occur (phase 0.906,0.977,1.023 and 1.093
respectively).  These relations are further complicated by the
presence of starspots which in general have the effect of altering the
depth of the eclipses and making the contact points asymmetrical.

SV Cam has been the subject of numerous photometric and spectroscopic
studies.  Despite this, there still does not yet exist a established
and reliable set of stellar parameters.  The various published
combinations of stellar parameters that can be found for SV Cam in the
literature are summarised in Table ~\ref{param}.  A discussion of the
individual merits of each set of binary system parameters is beyond
the scope of this paper.  Given the wide range of geometric parameters
available for SV Cam, it was necessary to determine a set of
parameters that would give an optimal fit to our lightcurve.

In fitting the photometric lightcurve of SV Cam we use the mass ratio
(q=0.641$\pm$0.007), semi-amplitude velocity (K$_1$=121.86$\pm$0.76,
K$_2$=190.17$\pm$1.73) and systematic radial velocity (-9.31$\pm$0.78)
of \citet{rucinski02} as their solution for SV Cam's spectroscopic
orbit is very well defined.

\subsection{Temperature}

\citet{jefferspc05} determined the temperature of the separated 
primary and secondary components of SV Cam using {\sc phoenix} 
\citep{allard00} model atmosphere spectra.  Model atmospheres, 
ranging in temperature from 5600\,K to 6500\,K in 100\,K steps, were
fitted to the spectrum of the primary and secondary stars using
$\chi^2$ minimisation.  The minimum $\chi^2$ value corresponds to a
temperature of 6013$\pm$19\,K and 4804$\pm$143\,K for the primary and
secondary stars respectively.  The primary star was then isolated and
fitted using the same method, where the minimum temperature is
6038$\pm$58\,K.  In this work we use the independently fitted primary
temperature as the error estimate is more reliable.

\subsection{Radii}

The radii for the primary and secondary stars were determined using
the robust grid search method.  This method uses a grid of radii as
input to DoTS ranging from 1.17 R$_\odot$ to 1.26 R$_\odot$ for the
primary star, and 0.73 R$_\odot$ to 0.82 R$_\odot$ for the secondary
star in 0.005 R$_\odot$ intervals.  For each grid point a model with
the specified set of geometric parameters was iterated 25 times to
obtain the lowest value of the reduced $\chi^2$.  The inclination was
not included as it scales as (sin i)$^{-1}$, implying a negligible
difference between inclinations of 90$^\circ$ and 85$^\circ$.

\begin{figure}
\psfig{file=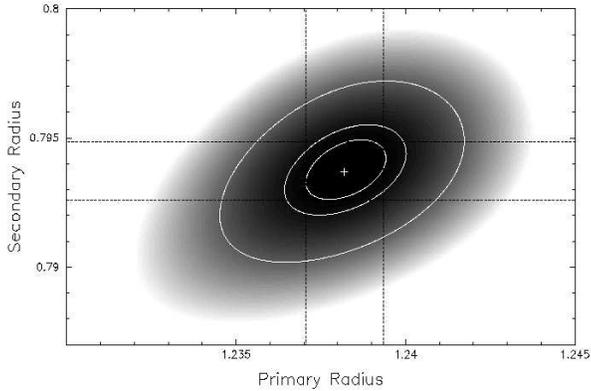,width=8.7cm,height=6.2cm,angle=270}
\caption{Contour plot of the $\chi^2$ landscape for the primary and
secondary radii.  From the centre of the plot, the first contour
ellipse is the 1 parameter 1$\sigma$ confidence limit at 63.8\%, the
second contour ellipse is the 2 parameter 1$\sigma$ confidence limit at
63.8\%, and the third contour ellipse is the 2 parameter 2.6$\sigma$
confidence limit at 99\%.}
\label{contrad}
\end{figure}

The results for the $\chi^2$ minimisation are shown in
Fig.~\ref{contrad} in the form of a contour plot.  The minimum
$\chi^2$ value occurs at 1.238\,$\pm$\,0.007 R$_\odot$ and
0.794\,$\pm$\,0.009 R$_\odot$ for the primary and secondary stars
respectively. The best fitting stellar parameters for the combined JGT
and HST data sets are summarised in Table ~\ref{geopar}.

\subsection{Lightcurve Fit and Surface Brightness Image}

\begin{figure*}
\hspace{-2cm}
\psfig{file=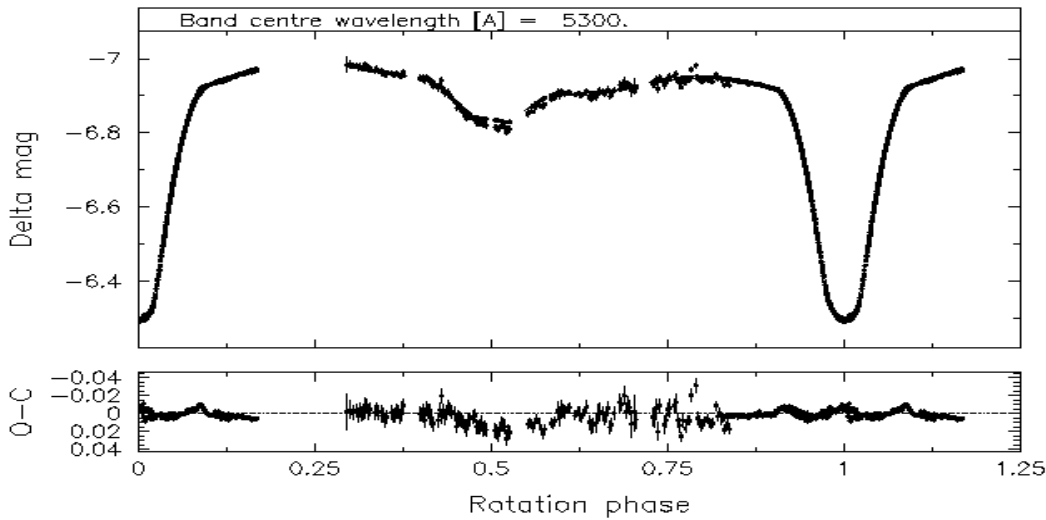,width=17cm,height=9cm,angle=270}
\vspace{-1.2cm}
\caption{Combined HST and JGT lightcurves, with a Maximum Entropy fit.}
%vspace{-3cm}
\label{fit}
\end{figure*}

\begin{figure}
\psfig{file=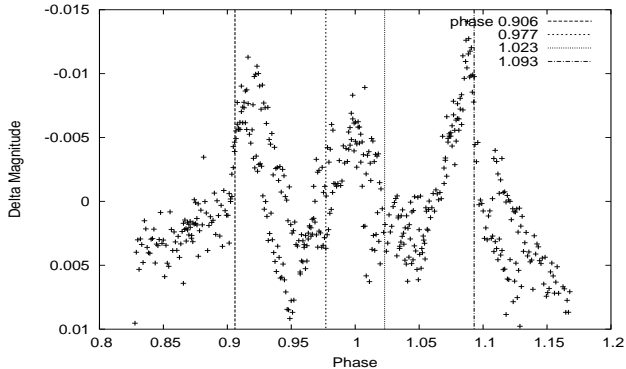,width=8.5cm,height=5cm,angle=270}
\caption{Observed minus Computed residuals with no polar cap (HST data
only).  The 
four contact points are shown as vertical lines.}
%\vspace{-0.4cm}
\label{o-c100}
\end{figure}

\begin{figure}
%\hspace{7cm}
\vspace{-0.5cm}
\psfig{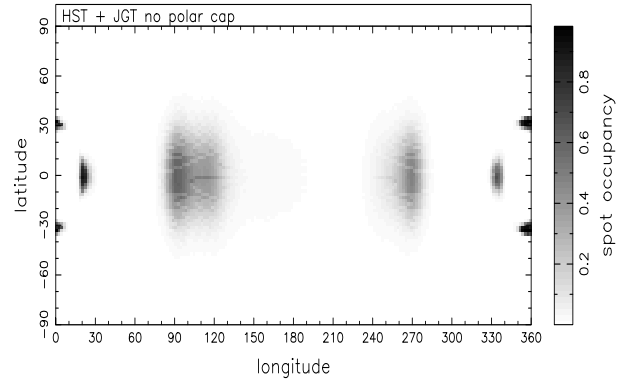}
\vspace{-3cm}
\caption{The final spot map of SV Cam (HST + JGT data).  Note that phase runs in reverse to longitude. Spots reconstructed using $\chi^2$ and located at the quadrature points have been shown by \citet{jeffersfs05} to be spurious.   Note that phase runs in reverse to 
longitude.}
%vspace{-3cm}
\label{spotmap}
\end{figure}

The resulting Maximum Entropy fit to the lightcurve, using the stellar
parameters as previously defined is shown in Fig.~\ref{fit}. The
observed data minus computed lightcurve residuals are shown in
Fig.~\ref{o-c100} in enlarged form covering just the HST primary
eclipse data.  The stellar surface image that results from these fits
is shown in Fig.~\ref{spotmap}.

The model lightcurve of the best fitting binary system parameters is
shown in Fig.~\ref{fit}.  When the modelled light-curve is subtracted
from the observed data the residuals are symmetric about phase 1, and
have two large discontinuities that occur at the first and fourth
contact points either side of the primary eclipse (Fig.~\ref{o-c100}).
Given the location of these discontinuities, an obvious explanation would be
that the radius of either of the two stars has been incorrectly
determined. However, the accuracy of our determination of the stellar
parameters, as described in Section 4, can eliminate this possibility.

The surface brightness distribution (Fig.~\ref{spotmap}) shows two
large spot features at the quadrature points similar to 'active
longitudes' that have been observed using photometry on other RS CVns
(e.g. \citet{olah91longitudes}, \citet{lanza01} and \citet{lanza02}).
These spots are reconstructed at the equator as there is no eclipse
information to determine the latitude.
The spot at 100$^\circ$ is stronger as that quadrature of the star has
a higher spot coverage, which is also evident in the lower light level
of the photometric lightcurve at phase $\approx$0.75. The spots
reconstructed at 25$^\circ$, 335$^\circ$, and 360$^\circ$ result from
spots on the primary star that have been occulted by the secondary
star during primary eclipse.  The symmetry is due to the inability 
of DoTS to determine the hemisphere of the spot feature.

\section{Unresolvable Spot Coverage}

\begin{table}

\protect\label{geopar}

\fontsize{6}{8}\selectfont  
\hspace{0.4cm}
\begin{tabular}{l l l}

%\begin{tabular}{ l l l l l l } 

\hline
\hline
Parameter & sol. 1 & sol. 2 \\

\hline

Ephemeris & 52214.34475 (MJD)\\
T$_{pri}$ & 6038\,$\pm$58 K &  5935\,$\pm$38 K \\
T$_{sec}$ & 4804\,$\pm$143 K & 4804\,$\pm$143 K \\ 
R$_{pri}$ & 1.238\,$\pm$0.006 R$_\odot$ & 1.235\,$\pm$0.007 R$_\odot$ \\
R$_{sec}$ & 0.794\,$\pm$0.006 R$_\odot$ & 0.727\,$\pm$0.006 R$_\odot$ \\
Polar Cap & & 46.7$^o$\\

\hline
\hline
\end{tabular}
\caption{Geometric binary system parameters computed for SV Cam.
For solution 1, the primary and secondary temperatures are taken from
\citet{jefferspc05}, while solution 2 shows the radii and polar cap size
solved using a reduced photospheric temperature to synthesise the star
being peppered with small spots.}
\end{table}

\subsection{Reduced Photospheric Temperature}

\citet{jeffersfs05} tested the limitations of surface brightness 
images from photometric data by modelling many sub-resolution spots on
an immaculate SV Cam.  Surface brightness distributions reconstructed
from these synthetic lightcurves show distinctive spots at the
quadrature points.  The presence of similar spot features in
Fig.~\ref{spotmap} could indicate that the surface of SV Cam is
peppered with many spots that are below the resolution capabilities of
the eclipse mapping technique.  This would then be consistent with the
results of TiO band monitoring studies \citep{oneal98tio} which show that
between 30\% and 50\% of a star's surface may be spotted.  

Further evidence for the peppering of the SV Cam's primary star by
small starspots is shown by \citet{jefferspc05}.  As previously
described in section 4, they determined the temperatures
of the two stars using best-fitting {\sc phoenix} model atmospheres.
It was also found that the primary's surface flux is approx 28\%
lower than predicted by a {\sc phoenix} model atmosphere at the best
fitting effective temperature.  Even taking into account the spot
distributions as shown in Fig.~\ref{spotmap}, this flux deficit can
only be accounted for if the primary star's surface is peppered with
unresolved spots.  As these spots are not resolvable using the
eclipse-mapping technique they can lead to a under estimation of the
star's photospheric temperature and will have the effect of
decreasing the flux deficit during the eclipse at all wavelengths.

\begin{figure}
\psfig{file=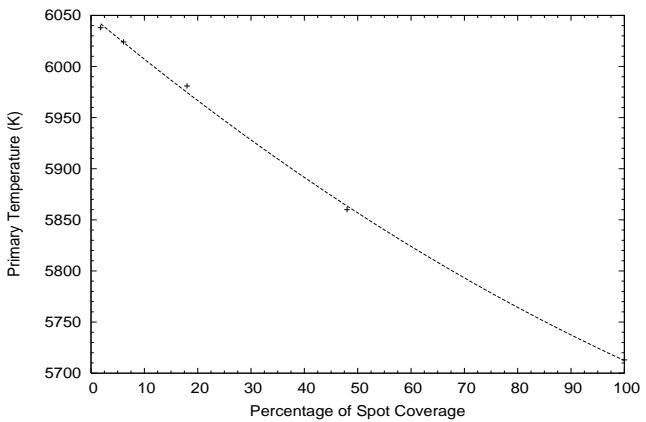,width=8.7cm,height=5.6cm,angle=270}
\caption{The decrease of primary star's photospheric temperature
as a function of percentage of spot coverage on its surface.}
\label{newtemp}
\end{figure}

To determine the reduced photospheric temperature, resulting from the
presence of many unresolvable spots, we extend the work of
\citet{jeffersfs05}.  An extrapolated solar spot size
distribution is applied to an immaculate SV Cam, for 1.8\%, 6.1\%,
18\%, 48\% and 100\% area filling factors of spots, with
T$_{ph}$=6038\,K, and T$_{sp}$=4538\,K.  Each of these spot
distributions is modelled as a photometric lightcurve, and is then
used as input to the Maximum Entropy eclipse mapping code.  To obtain
a satisfactory fit to the lightcurve it was necessary to reduce the
photospheric temperature to values as shown in Fig.~\ref{newtemp}.  A
quadratic fit to these points gives a temperature for 28\% spottedness
of 5935\,K which is the average temperature over hotter and
unresolvable cooler temperature regions.

\subsection{Polar Spot}

The work of \citet{jefferspc05} also showed that in addition to the 28\%
spot coverage in the eclipsed region of the primary star, there is an
additional 12.5\% flux deficit in non-eclipsed regions of the star.
The additional flux deficit can only be explained by the presence of a
polar cap which would have the effect of increasing the depth of the
primary eclipse, and would lead to an incorrect lightcurve solution.

\begin{figure}
\psfig{file=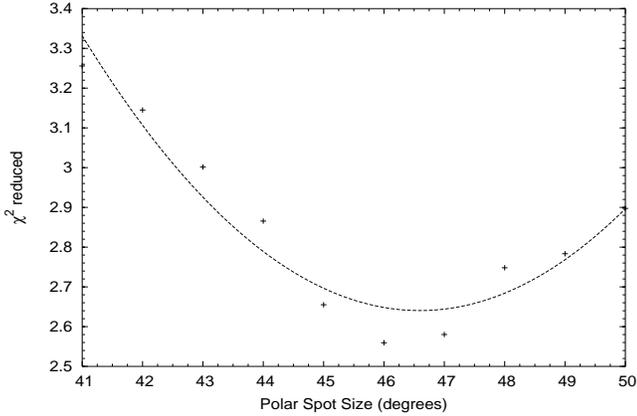,width=8.7cm,height=5.6cm,angle=270}
\caption{Quadratic fit to the variation of $\chi^2$ as a function of 
polar spot size.}
\label{chipcap}
\end{figure}

\begin{figure}
\psfig{file=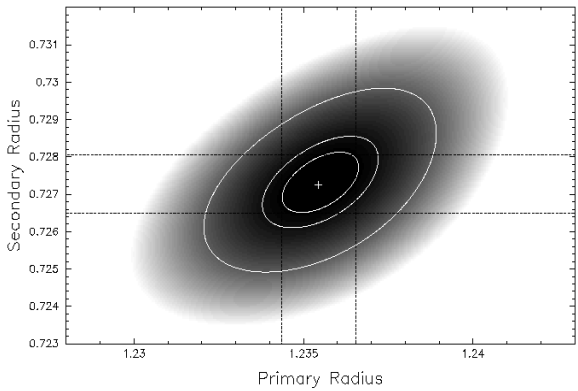,width=8.7cm,height=6.2cm,angle=270}
\caption{Contour plot of the $\chi^2$ landscape for the primary and
secondary radii with a 46.69$^\circ$ polar spot.  From the centre of
the plot, the first contour ellipse is the 1 parameter 1$\sigma$
confidence limit at 63.8\%, the second contour ellipse is the 2
parameter 1$\sigma$ confidence limit at 63.8\%, and the third contour
ellipse is the 2 parameter 2.6$\sigma$ confidence limit at 99\%.}
\label{pcaprad}
\end{figure}

We include a polar cap in our lightcurve solution by extending the
2-dimensional grid search described in section 4 to a 3-dimensional
grid search including the polar spot size from 35$^\circ$ to
50$^\circ$.  The polar spot is assumed to be at 4500\,K, circular,
centred at the pole, and is in addition to the reduced photospheric
temperature as described above.  For each polar spot size the minimum
primary and secondary radii are determined using a $\chi^2$ contour
map.  These minimum $\chi^2$ values are plotted as a function of polar
spot size in Fig.~\ref{chipcap}.  The best fitting polar cap size,
46.7$^\circ$ was determined from the minimum of a quadratic function
fitted to these points.  This value is in good agreement with the
independently determined value of 42$\pm$6$^\circ$ by
\citet{jefferspc05}.
The grid search of radii is then repeated using a fixed value for the polar
spot size.  The results for the $\chi^2$ minimisation are summarised
in Table 3 and are shown as a contour map in Fig.~\ref{pcaprad}.

\begin{figure*}
\hspace{-2cm}
\psfig{file=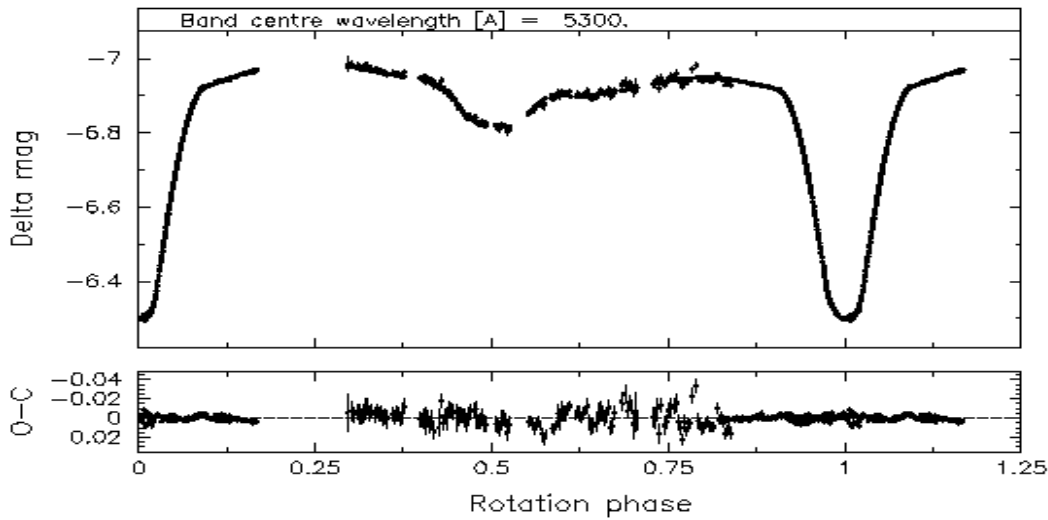,width=17cm,height=9cm,angle=270}
\vspace{-1.2cm}
\caption{Combined HST and JGT lightcurves, with a Maximum Entropy fit
with a polar cap and reduced photospheric temperature included.}
%vspace{-3cm}
\label{pcfit}
\end{figure*}

\begin{figure}
%\hspace{7cm}
\vspace{-0.5cm}
\psfig{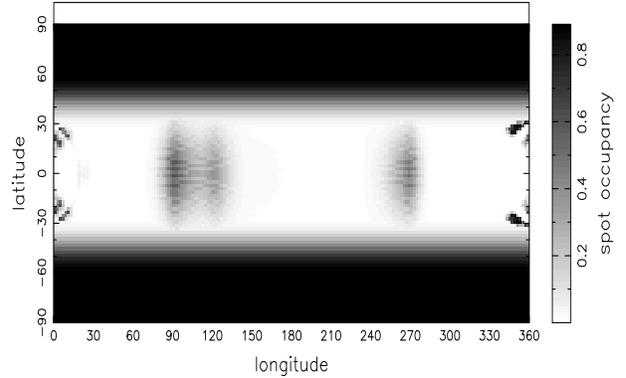}
\vspace{-3cm}
\caption{The final spot map of SV Cam (HST + JGT data) with two 46.7$^\circ$ 
polar caps and reduced photospheric temperature.  Note that phase runs
in reverse to longitude. Spots reconstructed using $\chi^2$ and
located at the quadrature points have been shown by \citet{jeffersfs05}
to be spurious as they can be reconstructed from models of high
starspot coverage.  Note that phase runs in reverse to longitude.}
%vspace{-3cm}
\label{pcspotmap}
\end{figure}

\begin{figure}
\psfig{file=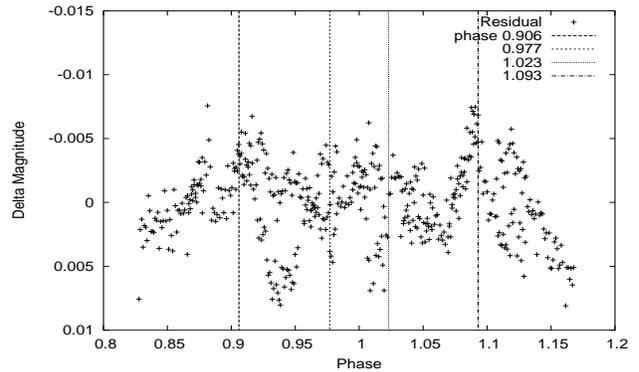,width=8.5cm,height=5cm,angle=270}
\caption{Observed minus Computed residuals for the case with a reduced 
photospheric temperature and a polar cap of 46.69$^\circ$ radius.  The 
four contact points are shown as vertical lines.}
%\vspace{-0.4cm}
\label{resnpc}
\end{figure}

\subsection{Final Lightcurve Fit}

The best fitting stellar parameters are used to create a Maximum 
Entropy lightcurve fit (Fig.~\ref{pcfit}) and stellar surface brightness
image for the primary star (Fig.~\ref{pcspotmap}). 
The observed data minus computed lightcurve residual for the primary
eclipse, as shown in Fig.~\ref{resnpc}, is significantly flatter than
the residuals shown in Fig.~\ref{o-c100}.  The secondary eclipse shows
also a better fit than in Fig.\ref{fit}.  The reduced $\chi^2$ value
after 25 iterations was 2.05, which is slightly reduced from the value
of 3.55 obtained without a polar spot and a reduced photospheric
temperature.  This result shows that the presence of a polar cap and
sub-resolution spots have a significant influence on the O-C
residuals.

The Maximum Entropy surface brightness distribution, as shown in
Fig.~\ref{pcspotmap}, is similar to that shown in Fig.~\ref{spotmap}.
The spot feature at 0$^\circ$ and 360$^\circ$ is not an artifact as it
is in the eclipse path of the secondary, but results from time
variable structure in the base of the primary eclipse.  It is a
symmetric spot feature as image reconstructions from photometry are
not able to determine a spot's latitude.  The spot feature at
25$^\circ$ and 335$^\circ$ in Fig.~\ref{spotmap} has disappeared due
to the correct determination of the primary star's temperature and
consequently the secondary star's radius.

\section{Discussion and Conclusions}

%notes: 
%* again independent conformation of polar caps on mag active stars.
%* solving of binary system parameters

We have used eclipse mapping, based on the maximum entropy method to
recover images of the visual surface brightness distribution of the
primary component of the RS CVn eclipsing binary SV Cam.  It is only
with the unprecedented photometric precision of the HST data that it is
possible to see strong discontinuities at the four contact points in the
residuals of the fit to the lightcurve.  These features can only be
removed from the O-C lightcurve by the reduction of the photospheric
temperature, to synthesise high unresolvable spot coverage, and the
inclusion of a polar spot.

In fitting SV Cam's lightcurve we used the independently determined
primary and secondary temperatures, as determined by
\citet{jefferspc05}, as input to our solution.  The primary and 
secondary radii were then determined using a `grid search' method.
Although the `grid search' method is computationally intensive, it
does provide reliable results.  In contrast to this is the `downhill
simplex' amoeba method \citep{press92} which repeatedly returned
lightcurve solutions from local $\chi^2$ minima rather than from the
global $\chi^2$ minimum.  The resulting parameters
(R$_{pri}$=1.238R$_\odot$ and R$_{sec}$=0.794R$_\odot$) are in closest
agreement with the values of \citet{pojmanski98} derived from
spectroscopic data (R$_{pri}$=1.25R$_\odot$ and
R$_{sec}$=0.8R$_\odot$).  The errors derived in this work are an order
of magnitude smaller than previous lightcurve solutions.

The Maximum Entropy fit to the data using the best-fitting orbital
parameters shows strong discontinuities at the four contact points of
the primary eclipse.  When we reconstructed an image using {\em only}
the ground based JGT data (photometric precision = 0.006) we found
that these strong discontinuities were not visible in the residual
lightcurve.  It is only with the high photometric precision of the HST
(0.00015 mag. or S/N 5000) that it is possible to distinguish the
strong discontinuities in the residual lightcurve.

The surface brightness distribution shows spots at the quadrature
points consistent with surface images of other RS CVns reconstructed
using $\chi^2$ minimisation e.g. \citet{olah91longitudes} and
\citet{lanza01}, \citet{lanza02}.  However, the reliability 
of these images has been investigated by \citet{jeffersfs05} where
spurious spots at the quadrature points have been reconstructed from
a synthetic star containing a high degree of sub-resolution spots.
This is consistent with results from other methods such as TiO-band
monitoring \citep{oneal98tio} which indicate that between 30\% and
50\% of a star's surface could be covered in starspots at all times.

A high total spot coverage on SV Cam's primary star will modify the
apparent photospheric temperature.  In a related paper
\citet{jefferspc05} showed that when the surface flux in the
low-latitude eclipsed region was approximately 30\% lower than the the
best fitting {\sc phoenix} model atmosphere.  In this paper we showed
that this is equivalent to a reduction in the photospheric temperature
from 6039$\pm$58\,K to 5935$\pm$38\,K.  High spot coverage also has
important structural implications.  As investigated by
\cite{spruit86spots}, a star with high spot coverage will, over
thermal timescales, readjust its structure to compensate in radius and
temperature.  For example the primary star's radius is 10\% larger
than expected for its spectral type, and could provide an explanation
as to the large variation in binary system parameters as shown in
Table~\ref{param}.  

The addition of a polar spot and the reduction of the photospheric
temperature has a negligible influence on the empirically determined
radius of the primary star, but a more significant influence on the
secondary, decreasing it from 0.794$\pm$0.007\,R$_\odot$ to
0.727$\pm$0.009\,R$_\odot$.  The presence of a polar spot will
increase the depth of the primary eclipse, while a reduction in the
photospheric temperature will decrease the depth of the eclipse.  In
both cases the binary system compensates for this by making the
secondary star smaller as it consequently needs to eclipse less light.
The standard luminosity-radius-temperature relation
(L=4$\pi$R$^2$$\sigma$T$^4$) shows that if T$_{pri}$ and T$_{sec}$ are
fixed the only variable parameters are the radii of the primary and
secondary stars.  The timing of the first and fourth contact points
fixes the primary radius, leaving the secondary radius as the only
adjustable parameter.

The resulting residuals from the Maximum Entropy lightcurve fit are
significantly flatter with the lightcurve solution that includes a
polar cap and a reduced photospheric temperature.  However, there are
still variations in the residual lightcurve, which could result from
the eclipse mapping routine interpreting small scale variations in the
lightcurve (i.e. unfitted spots) as noise.  A 0.005 change magnitude
as shown in Fig.~\ref{resnpc} would result in additional 14\% of the
area of the annulus around the secondary being spotted than is shown in
the surface brightness images.  However, the remaining structure in
the O-C residual could result from an incorrect value of the limb
darkening or the presence of plage on the primary star's surface,
which will require further investigation.

The surface brightness image reconstructed with the refined binary
system parameters (Fig.~\ref{pcspotmap}) has reduced surface structure
compared to the image reconstructed with the original parameters.
This is noticeable at longitudes 25$^\circ$ and 335$^\circ$, where the
small spots have disappeared with the refined binary system
parameters.  The structure at longitudes 0$^\circ$ and 360$^\circ$ are
not artifacts of the eclipse mapping technique as they are in the
eclipse path of the secondary, but result from time variable structure
on the surface of the primary star during primary eclipse.  The
features are symmetric as it is not possible to resolve a starspots
latitude using photometric observations.  Doppler images of active
stars show that polar caps do not have a uniform structure, unlike the
polar caps used in our model and shown in Fig.~\ref{pcspotmap}.

The upcoming COROT and KEPLER missions, which are designed to detect
transits eclipses of stars by terrestrial sized planets, will discover
thousands of eclipsing binary stars with micromag photometry, while
missions such as GAIA should deliver 10$^6$ new eclipsing binary
stars.  The benefit of such a wealth of high precision data will be
lost if it is not possible to accurately solve the binary system
parameters from the lightcurves.

\section*{Acknowledgments}

We would like to thank Keith Horne for the use of his XCAL program and
Roger Stapleton and Tim Lister for their guidance on the time
corrections.  We would like to also thank Phil Hodge at STSCI for
amending the HST data reduction pipeline routine `odelaytime', to
account for heliocentric and satellite-Earth time corrections.  SVJ
acknowledges support from a PPARC research studentship and a
scholarship from the University of St Andrews while at St Andrews
University.  
SVJ currently acknowledges support at OMP from a personal Marie Curie
Intra-European Fellowship within the 6th European Community Framework
Programme.

\bibliographystyle{mn2e}
\bibliography{iau_journals,master,ownrefs}

\bsp
\label{lastpage}

\end{document}